\documentclass[prl,aps,showpacs,superscriptaddress,twocolumn]{revtex4}
\usepackage{graphicx}
\usepackage{amsmath}
\usepackage{amsfonts}
\usepackage{amssymb}
\usepackage{epsfig}

\begin{document}
\title{
Scaling behavior of the adiabatic Dicke Model}
\author{Giuseppe Liberti}\email{liberti@fis.unical.it}\affiliation{
Dipartimento di Fisica, Universit\`a della Calabria, 87036
Arcavacata di Rende (CS) Italy}\affiliation{INFN - Gruppo
collegato di Cosenza, 87036 Arcavacata di Rende (CS) Italy}
\author{Francesco Plastina}\affiliation{ Dipartimento di Fisica, Universit\`a della
Calabria, 87036 Arcavacata di Rende (CS) Italy} \affiliation{INFN
- Gruppo collegato di Cosenza, 87036 Arcavacata di Rende (CS)
Italy}
\author{Franco Piperno}\affiliation{
Dipartimento di Fisica, Universit\`a della Calabria, 87036
Arcavacata di Rende (CS) Italy}
\date{\today}

\date{\today}

\begin{abstract}
We analyze the quantum phase transition for a set of $N$-two level
systems interacting with a bosonic mode in the adiabatic regime.
Through the Born-Oppenheimer approximation, we obtain the
finite-size scaling expansion for many physical observables and,
in particular, for the entanglement content of the system.
\end{abstract}
\bigskip
 \maketitle

\paragraph{Introduction.} Many-body systems have become of central
interest in the realm of quantum information theory as training
grounds to study static, dynamical and sharing properties of
quantum correlations. It has been natural, then, to employ the
entanglement as a tool to analyze quantum phase transitions, one
of the most striking consequences of quantum correlations in many
body systems \cite{nature,osborne,vida2}. In this letter, we study
the Dicke super-radiant phase transition and obtain the finite
size scaling behavior of the quantum correlations at criticality.

The Dicke model (DM) describes the interaction of $N$ two-level
systems (qubits) with a single bosonic mode \cite{dicke}, and  has
become a paradigmatic example of collective quantum behavior. It
exhibits a second-order phase transition \cite{hepp}, which has
been studied extensively \cite{WH,gilmore,liberti}. The continued
interest in the DM arises from its broad application range
\cite{phyrep} and from its rich dynamics, displaying many
non-classical features \cite{milburn,brandes,Hou,orszagent}. The
ground state entanglement of the DM has been recently analyzed
\cite{lambert,lambert2,reslen}, and some aspect of its finite size
behavior has been obtained numerically. Finally, Vidal and Dusuel,
\cite{vidal}, obtained the critical exponents by a modified
Holstein-Primakoff approach.

The exact treatment of the finite-size corrections to the Dicke
transition is quite complicated and the study of some limiting
cases can be useful. In this Letter we analyze the case of $N$
qubits coupled to a {\it slow} oscillator. We discuss on equal
footings both the finite size and the thermodynamic limit of the
model, thus allowing to obtain the phase transition as well as its
precursors at finite $N$. Indeed, we obtain the dominant scaling
behavior for some entanglement measures and the entire $1/N$
expansion for all of the relevant physical observables, such as
the order parameter. Concerning the quantum correlations, we argue
that what is really relevant for the critical behavior is the
bi-partite entanglement between the oscillator mode and the set of
two level systems.
\paragraph{Adiabatic limit.}\label{sec1}
The interaction of $N$ identical qubits with a bosonic mode is
described by the Hamiltonian
\begin{equation}
    H=\omega a^{\dagger}a+\Delta S_x+
   {\lambda}(a^{\dagger}+a)S_z/{\sqrt{N}}
    \label{1}
    \end{equation}
where $\omega$ is the frequency of the oscillator, $\Delta$ is the
transition frequency of the qubit and $\lambda$ is the coupling
strength. The $S$'s are the total spin observables,
${S}_{\nu}=\sum_{i=1}^{N}{\sigma}_{\nu}^{i}$, where
${\sigma}_{\nu}^{i}$ is the $\nu$-th the Pauli matrix for the
$i$-th qubit. $H$ is equivalent to the Dicke Hamiltonian
\cite{dicke}, obtained after the rotation $e^{i\pi S_y/4}$.

We assume a {\it slow} oscillator and work in the regime
$\Delta\gg\omega$ by employing the Born-Oppenheimer approximation.
As detailed in \cite{adiabatic}, the procedure can be followed
more plainly by rewriting the Hamiltonian of Eq. (\ref{1}) as
\begin{equation}
   H=\frac{\omega}{2}\left[P^2+Q^2+D S_x+\frac{L Q}{\sqrt{N}}  S_z\right] \, ,
    \label{hc}
\end{equation}
where we have introduced the dimensionless parameters
$D=2\Delta/\omega$, and $L=2\sqrt{2} \lambda / \omega$, together
with the oscillator coordinates $Q =(a^{\dagger}+a)/\sqrt{2}$, and
$P=i(a^{\dagger}-a)/\sqrt{2}$.

The basic assumption of the well-known adiabatic approximation is
that the state of a composite system with one fast and one slowly
changing part can be written as:
\begin{equation}\label{deco}
    |\psi_{tot}\rangle=\int d Q \, \phi(Q) |Q\rangle \otimes |\chi (Q)\rangle
\end{equation}
$|\chi (Q)\rangle$ is the eigenstate of the ``adiabatic'' qubit
equation for each fixed value of the slow variable $Q$,
\begin{equation}\label{adiaham}
    \left(D S_x+\frac{L Q}{\sqrt{N}}  S_z\right)|\chi(Q)\rangle=E(Q)|\chi(Q)\rangle \,,
\end{equation}
and can be written as the direct product of the eigenstates of
each single qubit
\begin{equation}\label{qubitstates}
    |\chi (Q)\rangle=|\chi (Q)\rangle_1\otimes|\chi
    (Q)\rangle_2\otimes\dots\otimes|\chi (Q)\rangle_N \, .
\end{equation}
As the qubits are identical, the lowest eigenstate of Eq.
(\ref{adiaham}) has the form
\begin{equation}\label{gsdgen}
|\chi_0(Q)\rangle = \left \{
\frac{1}{\sqrt{2}}\left[A_-(Q)|+\rangle-A_+(Q)|-\rangle\right]
\right \}^{\otimes N} ,
\end{equation}
where $|\pm\rangle$ are the $\pm 1$ eigenstates of $\sigma_z$, and
\begin{equation}\label{gsdgen2}
    A_{\pm}(Q)=\sqrt{{1\pm\frac{LQ}{\sqrt{N} \Theta(Q)}}} \, .
\end{equation}
The eigenvalue corresponding to this state is given by
\begin{equation}\label{dq}
E_0(Q)=-N\Theta(Q)= -N \sqrt{D^2+\frac{L^2Q^2}{N}}\,.
\end{equation}
This energy eigenvalue contributes an effective adiabatic
potential felt by the slow bosonic mode. The total potential for
the variable $Q$ is, therefore
\begin{equation}\label{udq}
    U_0(Q)=\frac{\omega}{2}\left[Q^2 - N \Theta(Q)\right] \, .
\end{equation}
Introducing the dimensionless parameter $\alpha={L^2}/{2 D}$, one
can show that for $\alpha\leq1$, the potential $U_{0}(Q)$ can be
viewed as a broadened harmonic potential well with its minimum at
$Q=0$. For $\alpha>1$, on the other hand, the coupling with the
qubit produces a symmetric double well with minima at $Q=\pm
Q_0=\pm{\sqrt{N}D}\sqrt{\alpha^2-1}/L$.
\paragraph{Ground state properties.}
In order to obtain the fundamental level of the coupled system,
the last step in the adiabatic procedure is the evaluation of the
ground state wave function for the oscillator, $\phi_0(Q)$, to be
inserted in Eq. (\ref{deco}). This wave function satisfies the
one-dimensional time independent Schr\"odinger equation
\begin{equation}\label{se}
\left(-\frac{\omega}{2}\frac{d^2}{dQ^2}+
U_{0}(Q)\right)\phi_{0}(Q)=\varepsilon_{0}\phi_{0}(Q) \, ,
\end{equation}
where $\varepsilon_{0}$ is the lowest eigenvalue.

The ground-state properties can be easily studied in the adiabatic
limit. It turns out that all of the expectation values can be
expressed in terms of the quantity:
\begin{equation}\label{genint}
   \Phi_\nu =\int_{-\infty}^\infty \phi_{0}^2(Q)
   \left({1+\frac{2\alpha}{ND}Q^2}\right)^{\nu}
   dQ\,,
\end{equation}
which is shown below to depend only on $\alpha$ and $ND$.

For the average values of the various components of the total
spin, one gets
\begin{equation}\label{magn}
   \frac{\langle S_x\rangle}{N}=-\Phi_{-\frac{1}{2}}\,,\quad
   {\langle S_y\rangle}={\langle S_z\rangle}=0\,,
\end{equation}
\begin{equation}\label{2pcfx}
    \frac{\langle
    S_x^2\rangle}{N^2}=\frac{1}{N}+\left(1-\frac{1}{N}\right)\Phi_{-1}\,,
\end{equation}
\begin{equation}\label{2pcfz}
    \frac{\langle S_z^2\rangle}{N^2}=\left(1+\frac{1}{N}\right)-\frac{\langle S_x^2\rangle}{N^2}
    \,,\quad    \frac{\langle S_y^2\rangle}{N^2}=\frac{1}{N}\, .
\end{equation}
Furthermore, the order parameter can be obtained from
\begin{equation}\label{orpa}
   \frac{\langle H\rangle}{N}=
   \frac{\omega}{2}\left(\frac{\langle{P^2+Q^2}\rangle}{N}
   - D\Phi_{\frac{1}{2}}(N)\right)\,.
\end{equation}

In the thermodynamic limit ($N\rightarrow\infty$) one gets the
well-known second-order quantum phase transition at
$\alpha=\alpha_c=1$, for which:
\begin{equation}
 \frac{\langle S_x\rangle}{N}=\left\{%
\begin{array}{ll}
    -1 & \hbox{$(\alpha\leq 1)$} \\
    \hbox{$-\frac{1}{\alpha}$} & \hbox{$(\alpha> 1),$} \\
\end{array}%
\right.
\end{equation}
\begin{equation} \frac{\langle S_x^2\rangle}{N^2}=\left\{%
\begin{array}{ll}
    1 & \hbox{$(\alpha\leq 1)$} \\
    \hbox{$\frac{1}{\alpha^2}$} & \hbox{$(\alpha> 1),$} \\
\end{array}%
\right.\,
 \frac{\langle S_z^2\rangle}{N^2}=1-\frac{\langle
 S_x^2\rangle}{N^2}\,,\quad \frac{\langle S_y^2\rangle}{N^2}=0\,,
\end{equation}
\begin{equation}
 \frac{\langle{Q^2+P^2}\rangle}{N}=\left\{%
\begin{array}{ll}
    0 & \hbox{$(\alpha\leq 1)$} \\
    \hbox{$\frac{D^2}{L^2}\left(\alpha^2-1\right)$} & \hbox{$(\alpha> 1),$} \\
\end{array}%
\right.
\end{equation}
and, finally
\begin{equation}
\frac{E_0}{N}=\left\{%
\begin{array}{ll}
    -D & \hbox{$(\alpha\leq 1)$} \\
    \hbox{$-\frac{D}{2}\left(\alpha+\frac{1}{\alpha}\right)$} & \hbox{$(\alpha> 1),$} \\
\end{array}%
\right.
\end{equation}
where $E_0=2\varepsilon_0/\omega$.
\begin{figure}
 \includegraphics{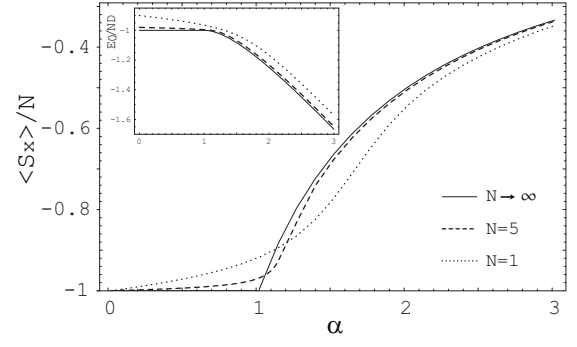}\\
 \caption{\label{envsa} The behavior of
$\langle S_x \rangle/N$ and of the ground state energy (inset) as
a function of the parameter $\alpha$, for $D=10$. Different curves
correspond to different values of $N$.}
\end{figure}
Numerical results for the ground state energy and the
$x$-magnetization are plotted in Fig.(\ref{envsa}) as a function
of the parameter $\alpha$, for different values of $N$, in
comparison with the results for $N\rightarrow\infty$.

In what follows, we will obtain an analytic expression for the $N$
dependence of $\Phi_{\nu} (ND)$, and, consequently, the scaling
relations for all the observable introduced above.
\paragraph{Finite-size scaling exponents at the critical point.}\label{sec2}
In the adiabatic regime, i.e. for large $D$, the Schr\"odinger
equation (\ref{se}) can be approximately rewritten as
\begin{eqnarray}\label{sescaling}
\left[-\frac{d^2}{dQ^2}+(\alpha_c-\alpha)Q^2+\frac{\alpha^2}{2ND}Q^4\right]\phi_{0}(Q;\alpha,ND)&&\nonumber\\
=e_{0}(\alpha,ND)\, \phi_{0}(Q;\alpha,ND)&&
\end{eqnarray}
with $e_{0}(\alpha,ND)=E_{0}(\alpha,ND)+ND$, \cite{quartic}.

Even thought the ground state energy and all of the physical
observables appear to depend on the two parameters $\alpha$ and
$ND$, this problem can be simplified in a single-parameter one
with the help of Symanzik scaling \cite{simon}, after transforming
Eq. (\ref{sescaling}) into the equivalent form
\begin{equation}\label{hamscaling}
\left[-\frac{d^2}{dq^2}+\zeta q^2+q^4\right]\phi_{0}(q;\zeta)
=e_{0}\left(\zeta\right)\phi_{0}(q;\zeta)
\end{equation}
where $q= Q \, \left (\frac{\alpha^2}{2 ND} \right )^{1/6}$, while
$\zeta=\left(\frac{2ND}{\alpha^2}\right)^{2/3}(\alpha_c-\alpha)$
is the only remaining scale parameter.

\noindent For the ground state energy one obtains the relation
\begin{equation}\label{sr}
E_{0}(\alpha,ND)=-ND +
\left(\frac{\alpha^2}{2ND}\right)^{1/3}e_{0}\left(\zeta\right) \,
.
\end{equation}
Taking the limit $\alpha\rightarrow\alpha_c$ (or $\zeta
\rightarrow 0$), Eq. (\ref{hamscaling}) becomes
\begin{equation}\label{pqo}
\left(-\frac{d^2}{dq^2}+q^4\right)\phi_{0}(q;0)=e_0(0)\phi_{0}(q;0)
\end{equation}
whose lowest eigenvalue is found to be $e_{0}(0)\simeq 1.06036$.
For $\zeta \ne 0$ (but near the critical point) we can resort to
perturbation theory and obtain the ground state energy as an
expansion in powers of $\zeta$,
\begin{equation}\label{ps}
e_{0}(\zeta)=\sum_{n=0}^\infty\beta_n\zeta^n\, .
\end{equation}
It is easy to show that $\beta_0= e_{0}(0)$ and
$\beta_1=\int_{-\infty}^\infty
q^2\phi_{0}^2(q;0)dq=e_0^\prime(0)\simeq 0.36203$. We demonstrate
below that these $\beta$'s enter not only the energy, but also the
finite $N$ expansion of every physical observable.

\begin{figure}
 \includegraphics{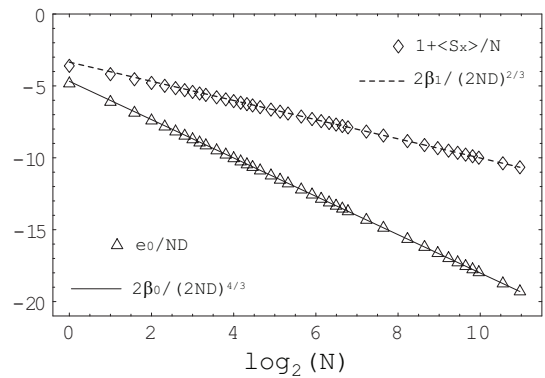}\\
 \caption{\label{enfn} Scaling of the ground state energy and spin average
 as a function of $N$ at the critical point $\alpha=1$, for $D=10$.}
\end{figure}

In order to compute the average values listed in Eqs.
(\ref{magn}-\ref{orpa}) at the critical point, we expand the
integral $\Phi_{\nu}(ND)$, given in Eq.(\ref{genint}), as
\begin{equation}\label{genint2}
   \Phi_\nu(ND) \simeq 1+\frac{2\alpha\nu}{ND} \langle
   Q^2 \rangle + \frac{2\alpha^2(\nu-1)\nu}{N^2D^2} \langle
   Q^4\rangle
\end{equation}
Taking the system energy $E_0(\alpha,ND)$ from Eq. (\ref{sr}), we
exploit the Feynman-Hellman theorem to obtain
\begin{eqnarray}\label{fh1}
&& \frac{\partial E_0(\alpha,ND)}{\partial \alpha} = -\langle Q^2
\rangle + \frac{\alpha}{ND}\langle Q^4 \rangle \, , \\ \label{fh2}
&& \frac{\partial E_0(\alpha,ND)}{\partial
(ND)}=-1-\frac{\alpha^2}{2N^2D^2} \langle Q^4\rangle \, .
\end{eqnarray}
Using Eq. (\ref{ps}), we get the critical point values
\begin{equation}\label{dm2}
\langle Q^2 \rangle ={\beta_1}\left(2ND\right)^{1/3} \, , \quad
\langle Q^4 \rangle =\frac{\beta_0}{3}\left(2ND\right)^{2/3} \,
\end{equation}
\begin{equation}\label{genint3}
   \Phi_\nu(ND)\simeq1+\frac{4\nu}{(2ND)^{2/3}}\beta_1+\frac{8}{3}\frac{(\nu-1)\nu}{(2ND)^{4/3}}\beta_0
   \, .
\end{equation}
Substituting into Eqs. (\ref{magn})-(\ref{2pcfx}), we have:
\begin{eqnarray}
&& \label{sxsca} \frac{\langle
S_x\rangle}{N}\simeq-1+\frac{2\beta_1}{(2ND)^{2/3}}-\frac{2\beta_0}{(2ND)^{4/3}}
\\
&& \label{sx2sca}
 \frac{\langle
    S_x^2\rangle}{N^2} \simeq  1-\frac{N-1}{N}\left[\frac{4\beta_1}{(2ND)^{2/3}}
     - \frac{16}{3}\frac{\beta_0}{(2ND)^{4/3}}\right]\,.
\end{eqnarray}
Finally, the order parameter is given by
\begin{equation}\label{posca}
  \frac{\langle{P^2+Q^2}\rangle}{ND}\simeq \frac{2\beta_1}{(2ND)^{2/3}}+\frac{4}{3}\frac{\beta_0}{(2ND)^{4/3}}\,.
\end{equation}
In Fig. (\ref{enfn}) we make a comparison with the results
obtained from the numerical solution of the Schr\"odinger equation
(\ref{se}). One can see that the leading finite size-corrections
for $e_0/ND$ and ${\langle S_x\rangle}/{N}$ scale indeed as
$N^{-4/3}$ and $N^{-2/3}$, respectively.
%
When the system size is too small, the next to leading orders in
these expansions become important and this explains the
discrepancy with Ref.\cite{reslen}, already pointed out in
\cite{vidal}.

\noindent The knowledge of $\beta_0$ and of the first two moments
is sufficient to determine recursively all the others ($k\geq 2$):
\begin{equation}\label{recrel}
  \frac{\langle Q^{k+4}\rangle}{(2ND)^{2/3}}=\frac{k+1}{k+3}\beta_0\langle Q^{k}\rangle+
  \frac{k(k^2-1)}{4(k+3)}(2ND)^{1/3}\langle Q^{k+2}\rangle.
\end{equation}
This relation allows one to compute higher order terms of the
finite size expansion.

We conclude this part by pointing out that the oscillator state
$\phi_0$ could also be found by a different re-scaling procedure,
i.e. by rewriting the Hamiltonian as $H(\xi)=p^2+ q^2+ \xi q^4$
with $\xi=\alpha^2/2ND(\alpha_c-\alpha)^{3/2}$. In this case,
however, the perturbation expansion would yield a power series
that diverges very strongly for every $\xi\neq0$. This would be
the equivalent in our approach of the method of Ref.\cite{vidal},
where the scaling exponents are obtained by arguing that there can
be no singularity in any physical quantity at finite-size. Our
leading order results agree with those reported in
Ref.\cite{vidal} except for the exponent of ${\langle
S_y^2\rangle}/{N^2}$, on which we comment below.
\paragraph{Entanglement.}\label{sec3}
To start the discussion on quantum correlations in this model, we
point out the peculiar nature of the adiabatic state of Eq.
(\ref{deco}); namely, the fact that, once the oscillator is traced
out, the $N$ qubits remain in a purely statistical mixture,
without the presence of any entanglement, neither of pairwise, nor
of multi-partite nature. In particular, the concurrence between
any two qubits is zero. This is due to the adiabatic hypothesis
$D\gg 1$, which strongly suppresses the energy exchange between
qubits, mediated by the oscillator.

Since we obtain all the relevant features of the Dicke transition
despite the fact that concurrence is neglected, we can conclude
that its presence is not really essential to describe the large
$N$ behavior. In fact, as reported by Ref. \cite{lambert2}, some
degree of pairwise entanglement is present even in the regime $D
\gg 1$ and therefore one should expect that the adiabatic approach
fails in describing pairwise correlations. Interestingly enough,
this is not entirely the case as, for large $N$, the only
two-point correlation function for which we obtain a different
scaling behavior is $\langle S_y^2\rangle$, and this difference is
just enough to set the concurrence to zero.

As argued above, the quantum correlations that really matter for
the phase transition are those involving the oscillator. Below, we
evaluate {\it i}) the entanglement of each qubit with the rest of
the system (as there is no entanglement among qubits, this is due
to correlations with the boson mode), and {\it ii}) the amount of
entanglement between the oscillator and the entire set of qubits.

\noindent After tracing out all of the other degrees of freedom,
each qubit is found in the same state $\rho_{1}$, and participates
of the entanglement (as measured by the tangle)
\begin{equation}\label{tau2}
\tau_1= 2 \Bigl [1 -\mbox{Tr} \rho_1^2 \Bigr ] \equiv
1-\frac{\langle S_x\rangle^2}{N^2} \, .
\end{equation}
In the thermodynamic limit, this is zero in the normal phase,
while one gets $\tau_1 = 1- \frac{1}{\alpha^2}$ for $\alpha > 1$.
For large but finite $N$, $\tau_1$ is non-zero even for $\alpha
\leq 1$; but at the critical point it scales as $\tau_1 \simeq
\frac{4 \beta_1}{(2ND)^{2/3}}$.
\begin{figure}
 \includegraphics{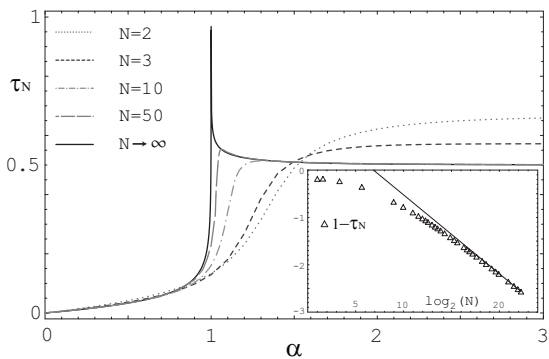}\\
 \caption{\label{tanglefiN} The tangle $\tau_N$ between the oscillator
 and the $N$ qubits as a function of $\alpha$, for $D=10$ and different
values of $N$. The inset shows the scaling of $\tau_N$  with $N$
at $\alpha= \alpha_c = 1$, compared with the analytic expression
of Eq. (\ref{scalat}) (continuous plot). }
\end{figure}

We adopt the linear entropy also to quantify the entanglement
between the oscillator and the $N$ qubits. Thus
\begin{equation}\label{line}
\tau_N = \eta \Bigl ( 1- \mbox{Tr} \left \{ \rho_{N}^2 \right \}
\Bigr ) \, , \end{equation} where $\rho_{N}= \mbox{Tr}_{osc} |\psi
\rangle \langle \psi|$ is the reduced qubit state, while the
pre-factor is chosen to be $\eta = \frac{2^N}{2^N-1}$ to  bound
$\tau_N$ to $1$, see Ref. \cite{scott}. Explicitly, we have
\begin{equation}
\mbox{Tr} \rho_{N}^2 = \int dQ d Q' \, \phi_0^2(Q) \phi_0^2(Q')
\Bigl [1+ \frac{\Theta^2(\sqrt{Q Q'})}{\Theta(Q) \, \Theta(Q')}
\Bigr]^N
\end{equation}
which can be expressed as a power series in $\Phi_{\nu}$.

Fig. (\ref{tanglefiN}) shows $\tau_N$ both for finite $N$ and for
$N \rightarrow \infty$. In the thermodynamic limit, we find
\begin{equation}
\tau_{\infty} =\left\{%
\begin{array}{ll}
   1- \left (1+ \frac{\alpha}{D \sqrt{1-\alpha}} \right )^{-\frac{1}{2}} & \hbox{$(\alpha\leq 1)$} \\
   1-\frac{1}{2} \left (1+ \frac{1}{D \alpha^2\sqrt{\alpha^2-1}} \right )^{-\frac{1}{2}} & \hbox{$(\alpha> 1),$} \\
\end{array}%
\right. \, ,
\end{equation}
which shows a cusp at the critical point, where $\tau_{\infty}
=1$. For finite $N$, this singular behavior is rounded and $\tau$
is quenched. When $N$ is large, the entanglement scales as
\begin{equation}
\tau_N(\alpha=\alpha_c)  \sim 1-  \,
\frac{\sqrt{\pi}\,K}{(2D)^{1/3}N^{1/6}} \, ,
\label{scalat}\end{equation} where $K= \int dq \phi_0^4(q;0)
\simeq 0.46$, and $\phi_0(q;0)$ is the normalized solution of Eq.
(\ref{pqo}). This result is shown in Fig. (\ref{tanglefiN}) to
agree with the numerical evaluation of Eq. (\ref{line}).

To summarize, we have described the finite size scaling of quantum
correlations in the Dicke model, obtaining the finite $N$ behavior
of the spin components, of the order parameter and of the ground
state entanglement in the adiabatic regime $D \gg 1$. We also
discussed the crucial role of the entanglement involving the
oscillator at criticality, giving its expression both in the
thermodynamic limit and for a finite size system.
%

\end{document}